\documentclass[conference]{IEEEtran}
\IEEEoverridecommandlockouts
\usepackage{cite}
\usepackage{amsmath,amssymb,amsfonts}
\usepackage{algorithmic}
\usepackage{graphicx}
\usepackage{courier}
\usepackage{pifont}
\usepackage{censor}
\usepackage{booktabs}
\usepackage{hyperref}
\usepackage{multirow}
\usepackage{textcomp}
\usepackage{url}
\usepackage[table]{xcolor}

\usepackage{array}
\usepackage{multirow}
\usepackage{geometry}
\usepackage{paralist}

\usepackage{placeins}

\def\BibTeX{{\rm B\kern-.05em{\sc i\kern-.025em b}\kern-.08em
    T\kern-.1667em\lower.7ex\hbox{E}\kern-.125emX}}
\begin{document}

\definecolor{strong_purple}{RGB}{97, 75, 195}
\definecolor{strong_green}{RGB}{51, 187, 197}
\definecolor{medium_green}{RGB}{133, 230, 197}
\definecolor{light_green}{RGB}{200, 255, 224}

\makeatletter
    \newcommand{\linebreakand}{%
      \end{@IEEEauthorhalign}
      \hfill\mbox{}\par
      \mbox{}\hfill\begin{@IEEEauthorhalign}
    }
    \makeatother

\title{Governance of Ledger-Anchored \\ Decentralized Identifiers \\
}

\author{\IEEEauthorblockN{Sandro Rodriguez Garzon, Carlo Segat, Axel Küpper}
\IEEEauthorblockA{\textit{Service-centric Networking} \\
\textit{Technische Universität Berlin / T-Labs}\\
Berlin, Germany \\
\{sandro.rodriguezgarzon\}\textbar\{carlo.segat\}\textbar\{axel.kuepper\}@tu-berlin.de}
}

\newcommand{\ts}{\textsuperscript}

\maketitle

\begin{abstract}

A Decentralized Identifier (DID) empowers an entity to prove control over a unique and self-issued identifier without relying on any identity provider. The public key material for the proof is encoded into an associated DID document (DDO). This is preferable shared via a distributed ledger because it guarantees algorithmically that everyone has access to the latest state of any tamper-proof DDO but only the entities in control of a DID are able to update theirs. Yet, it is possible to grant deputies the authority to update the DDO on behalf of the DID owner. However, the DID specification leaves largely open on how authorizations over a DDO are managed and enforced among multiple deputies. This article investigates what it means to govern a DID and discusses various forms of how a DID can be controlled by potentially more than one entity. It also presents a prototype of a DID-conform identifier management system where a selected set of governance policies are deployed as Smart Contracts. The article highlights the critical role of governance for the trustworthy and flexible deployment of ledger-anchored DIDs across various domains.

\end{abstract}

\begin{IEEEkeywords}
Decentralized Identifier, Digital Identity, Authorization, Governance, Blockchain, Distributed Ledger, Security, Trust\end{IEEEkeywords}

\section{Introduction}

The Self-sovereign Identity (SSI) paradigm \cite{Toth.2019} proclaims a new privacy-preserving way to handle digital identities. With SSI, individuals gain full control over their digital identity, deciding independently where the sensitive artifacts of their digital identity are stored and with whom, when, and for what purpose they are shared, all without requiring 3\textsuperscript{rd}-party permission. This stands in sharp contrast to traditional identity management where digital identities are centrally controlled by a few organizations and enterprises \cite{Avellaneda.2019}. The technical realization of SSI hinges on the ability of individuals to technically prove the validity of the self-controlled digital identity, including its identifier. To facilitate this, the W3C introduced the concept of Decentralized Identifiers (DIDs) \cite{WorldWideWebConsortium.822021}. A DID is a unique, self-issued identifier that refers to an entity, called \textit{DID subject}, such as an individual or a non-human entity. It points to the public key material required for others to verify a DID subject's proof of control over the DID. The private key material remains with the DID subject. When the DID's public key material is anchored in a publicly accessible verifiable data registry (VDR), such as a public distributed ledger, any relying party can independently and trustfully verify a proof of control over a self-issued DID presented by a DID subject. This eliminates the need for a 3\textsuperscript{rd}-party guardian, such as a certificate authority in public key infrastructures, to vouch for the DID's binding with its public key material. 

With a ledger-anchored DID, identity claims such as the DID subject's name or eye color can be cryptographically linked to the DID, be signed by trusted issuers, and handed over off-chain to the DID subject in form of Verifiable Credentials (VCs) \cite{WorldWideWebConsortium.VC} for free use. This empowers DID subjects to present tamper-proof VCs, for which they can prove to control the cryptographically linked DIDs, to relying parties also known as verifiers, without notifying or asking the issuers for permission. As a result, DIDs and VCs have become a cornerstone of self-sovereign digital identities, enabling individuals to govern and use their digital identities on its own and in a privacy-preserving manner rather than relying on identity providers.

According to the core principles of SSI \cite{Allen.2016}, any authoritative action related to a human-centered digital identity should be carried out by the very same individual. For example, a DID subject binding a new public key to their DID. But in practice and in particular with device IDs, a DID often needs to be controlled by others, even if the DID refers to a human being. For example, parents wish to control their minor children's DID, or the head of human resources wishes to control the DIDs of the company's employees \cite{did_use_cases}. To enable this, authority over a DID can be shared with or delegated to others. This article explores the different concepts and realizations for governing authority over a ledger-anchored DID. It investigates how a DID subject appoints deputies, what control rights a deputy possesses over a DID, and how potentially multiple deputies coordinate control of a commonly governed DID among each other. It also presents an exemplary DID-conform identifier management system realized using Smart Contracts (SCs) on Ethereum. The prototype permits to appoint multiple deputies with different authorizations for a DID, and its governance of DIDs is transparent and fully customizable via SCs and extensible and interchangeable during operation.  

The article starts in Section \ref{sec:did} with a brief introduction of the DID concept with a focus on governance and discusses in Section \ref{sec:related} governance aspects of existing DID methods and a few research approaches. Section \ref{sec:concept} investigates in detail what it means to govern a DID and lists potential improvements. Section \ref{sec:implementation} describes briefly a prototypical identifier management system that was built based on the findings in Section \ref{sec:concept}. The article concludes in Section \ref{sec:conclusion} with a short summary of the observations and sketches open research questions. 

\begin{figure*}[t!]
\centerline{\includegraphics[scale=0.9]{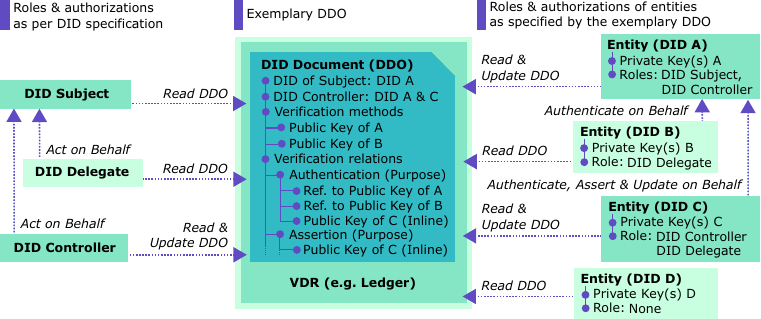}}
\caption{Exemplary schematic DDO with roles and authorizations.}
\label{fig:dids}
\end{figure*}

\section{Decentralized Identifier}
\label{sec:did}
A DID is a self-issued and unique identifier of the form {\texttt{did:[DID method]:[DID method-specific identifier]}}, for example, {\texttt{did:eth:123456}}. It makes the DID subject uniquely identifiable by the did method-specific identifier. A DID is resolvable and points to a DID document (DDO). The DDO contains besides meta information the public key material associated with the DID. The DDO enables the DID subject together with the securely kept private key material to cryptographically prove control over the DID.

Since the DID is self-issued it needs to be guaranteed that the DDO can be obtained by others in a tamper-proof way. Otherwise, a malicious actor would be able to replace the public key material of a DID subject's DDO by its own to impersonate the DID subject. A DDO is therefore preferably stored in a commonly accessible VDR in form of a trusted centralized database, a decentralized file system, or a distributed ledger. The DID method specifies how to securely create, read, update, and delete DDO's in a specific VDR. Since a distributed ledger guarantees algorithmically that a verifiable and up-to-date DDO can be read by all entities with access to the ledger but only the DID subject is able to modify theirs in the ledger, it became the first choice technology for anchoring DDOs.      

The way a DDO materializes itself in a ledger depends upon the DID method and ledger technology in use. Logically, the DDO contains verification methods - mostly public keys - for different purposes such as for authentication and assertion. It is possible to add verification methods of associated entities, called \textit{DID delegates}, so that they can, e.g., authenticate on behalf of the DID subject without holding the DID subject's private key material. DID delegates, however, are not authorized to update a DDO and are thus not involved in the governance of a DID. Hence, they can't prove to control the DID but to be granted the right - via the enlisting of its verification material - to represent the DID subject for a specific purpose. It is also permitted to delegate update rights for the DDO to other parties by declaring the DIDs of the deputies in the DDO. These deputies are called \textit{DID controllers}. They govern the DID through their delegated control of the DDO. Therefore, the DDO does not only specify the actions selected entities are authorized to perform on behalf of the DID subject, here named \textit{functional policies}, but also whether they are authorized to update the DID subject's DDO itself, here named \textit{governance policies}. The DID specification follows a role-based access control model \cite{Sandhu.1996} with a default-deny permission pattern. Authorizations in the DDO can only be of permissive and not of restrictive nature. If no explicit authorization is granted in the DDO, it will not be granted implicitly. A DID subject has no special role regarding authorizations and as such no implicit DDO update right. Hence, a DID subject is not in control of its own DDO unless it is explicitly assigned the role of a DID controller in the DDO. Figure \ref{fig:dids} illustrates schematically an exemplary DDO and the associated roles and authorizations. 

The basic permissions required to interact with a VDR in form of a ledger, such as the right to read a DDO from and write it into the ledger, are not declared in a DDO as they are enforced globally by the ledger. A ledger applies its own policies to govern the usage of its decentrally managed resources. However, if a ledger is of permissioned nature, then its access control layer needs to be aware of the update authorizations granted to an entity in the DDO on the application layer to avoid potential policy conflicts. For example, a DID subject might appoint a DID controller, but the DID controller's digital identity used to access the ledger is not granted write rights. To the best of our knowledge, Hyperledger Indy\footnote{\url{https://www.lfdecentralizedtrust.org/projects/hyperledger-indy}} is the only permissioned ledger technology that makes natively use of anchored DIDs at the application layer to assign authorizations on the ledger's access control layer. However, the assigned authorizations are not declared in the DDO and are highly specific to Hyperledger Indy.   

\begin{table*}[!ht]
\centering
\begin{tabular}{|>{\centering\arraybackslash}p{1.1cm}|>{\centering\arraybackslash}p{1.2cm}|>{\centering\arraybackslash}p{1.3cm}|>{\centering\arraybackslash}p{1.4cm}|p{7.8cm}|}
\hline
\rowcolor{strong_green} \textbf{DID method} & \textbf{Ledger} & \textbf{Controller  updatable} & \textbf{Controller in DDO} & \textbf{Control}\\ 

\hline

\rowcolor{light_green} did:algo & Algorand & \ding{51} & - & DID is fully controlled by an single Algorand SC. The latter can be controlled by multiple Algorand identities using a multisign wallet. Controllers are defined in the contract code.\\

did:bnb & Binance &  - & n/a & DID is fully controlled by a single Binance ID subject. Delegates can be appointed by Binance ID subject to add their verification methods to the DID. \\

\rowcolor{light_green} did:btcr & Bitcoin & - & - & DID is fully controlled by a single entity via its single key used for the latest DID-related Bitcoin transaction.  \\

did:ebsi (legal) & EBSI Blockchain & \ding{51} & \ding{51} & DID can be controlled by one or more EBSI entities, but unclear if DID controllers act independently or in a coordinated manner. \\

\rowcolor{light_green} did:sol & Solana & \ding{51} & \ding{51}* &  DID can be fully controlled by mutliple Solana accounts. *Controller property not used. Instead, keys of DID controllers listed in "capabilityInvocation" section of DDO. \\

did:ens & Ethereum & \ding{51} & - & DID is fully controlled by a single ETH account. Delegation explicitly allowed. Controller property not used. \\

\rowcolor{light_green} did:erc725 & Ethereum & \ding{51} & - & DID is fully controlled by a single ERC725 identity. An ERC725 identity can be controlled by multiple ETH accounts and changes of theses are permitted. \\

did:ethr & Ethereum & \ding{51} & - & DID is fully controlled by a single externally owned ETH account. A SC ETH account can be used instead to give more than one externally owned ETH account control power. \\

\rowcolor{light_green} did:factom & Factom & \ding{51} & - & DID can be controlled by multiple Factom entities. The extend to what a Factom entity in the role of a DID controller is permitted to change in a DDO depends upon its DID-specific priority. \\

did:icon & Ledger-agnostic & \ding{51} & n/a & DID is fully controlled by a single ICON identity. \\

\rowcolor{light_green} did:indy/ did:sov & Hyperledger Indy& \ding{51} & - & DID is fully controlled by a single legal Indy entity. \\

did:ion & Bitcoin & - & n/a & DID is fully controlled by a single Sidetree identity owner \\

\rowcolor{light_green} did:iota  & IOTA & \ding{51}  & \ding{51} &  DID can be fully controlled by multiple entities. One controller makes an update proposal, others approve it, and the update is executed as soon as a predefined threshold is reached.\\

did:jolo & Ethereum & - & n/a & DID is fully controlled by a single ETH account. \\

\rowcolor{light_green} did:orb & Sidetree Protocol &  n/a & n/a & DID is fully controlled by a single did:orb. Multiple controllers not supported by Sidetree protocol. Unclear if delegation is allowed.\\

did:stack & Ledger-agnostic & \ding{51} & n/a & DID is fully controlled by a single address of the underlying blockchain technology. \\

\hline
\end{tabular}
\vspace{0.2cm}
\caption{Ledger-based DID methods and their DID update governance}
\label{table:methods}
\vspace{-0.5cm}
\end{table*}

\section{Related Work}
\label{sec:related}
While the design and realization of the governance is crucial for the trustworthiness of decentralized identity management with DIDs, this topic has so far been only marginally addressed in research \cite{vulnerability_did_doc_update, modid, dad_3_rs}. On the other hand, numerous DID methods and their realizations already exist \cite{didmethods}, with each implementing their own governance. Comprehensive comparisons and classifications of DID methods are given by Hoops et al. in \cite{Hoops.2023} and Bistarelli et al. in \cite{Bistarelli.2023}, but they lack a discussion of the governance aspects. An overview of a selected set of popular ledger-based DID methods and their governance peculiarities are given is Table \ref{table:methods}. 

Most DID methods do not require a declaration of DID controllers in the DDO. Instead, they rely on VDR-technology-specific mechanisms to manage authorizations and to check whether an entity trying to update a DDO in the VDR is authorized to do so. It is common to grant only a single entity DDO update rights. Prominent examples are \texttt{did:ethr}\cite{didethr}, \texttt{did:bnb}\cite{didbnb}, \texttt{did:indy}\cite{didindy}, and \texttt{did:sov}\cite{didsov}. However, this is a rather inflexible model as it limits the application of DIDs to domains where a digital identity is only controlled by a single entity. Only a few DID methods support multiple DID controllers, e.g., \texttt{did:algo}\cite{didalgo}, \texttt{did:sol}\cite{didsol}, and \texttt{did:ebsi}\cite{didebsi}, but each DID controller is granted full DDO update rights. This does not follow the principle of least privilege as in certain cases a deputy is assigned more DDO update power than actually needed. A DID subject, on the other hand, might trust deputies to update their own verification methods in the DID subject's DDO after a key rotation but not to appoint additional deputies that the DID subject might not know. More fine-granular authorizations can be defined in \texttt{did:factom}\cite{didfactom} by giving each DID controller a priority. Such a DID controller is only permitted to update verification methods of a DID controller with the same or lower priority. In \texttt{did:iota}\cite{didiota}, a controller is able to make an update proposal. Other controllers of the same DDO approve the proposal. It is accepted and the update is executed as soon as a voting threshold is reached.   

Rhie et al. in \cite{vulnerability_did_doc_update} and Mazzocca et al. in \cite{Mazzocca.2025} point out that a misbehaving DID controller could potentially revoke the DID subject's control rights by means of a DDO update to gain sole control over the DID. Rhie et al. propose more fine-granular delegation policies to limit the DID controller's rights, but they do not state where these are declared and how they are enforced. Same applies to a proposed immutable invocation history, which would track all DID controller's off-ledger actions that happen on behalf of the DID subject. Yet, this approach stands in direct contradiction to the privacy-preserving core principles of SSI. Yang et al. realizes fine-granular delegation policies by the introduction of hierarchical identity management for "DID owners" (not to be mistaken with the DID subject) with the roles "root owner", "delegate", and "basic owner" \cite{modid}. A root owner is allowed, besides others, to add and remove root owners, a delegate is only involved in key recovery, and a basic owner corresponds to a DID representative. If multiple root owners are declared for a DID, then either one can add and remove root owners on its own without involving the others or an n-out-of-m consensus upon a group change request needs to be reached among the root owners. The approach was implemented with SCs in Ethereum. While it is a promising approach, they do not discuss whether role definitions and assignments need to be made explicit in the DDO. In \cite{dad_3_rs}, Smith addresses key rotation in a DDO under multiple controllers. The proposed method requires each party to register two public keys in the DDO: the active operational and a pre-rotated key. During key rotation, parties submit a transaction that includes signatures from both keys. As a result, the pre-rotated key becomes the new active operational key, and a new pre-rotated key is registered for future rotations. The relevance of Smith's work lies in it's partial addressing of coordinated DDO updates. However, an in-depth investigation into ledger-anchored DDOs' governance remains absent, with aspects such as delegation and enforcement not being addressed.

\section{Governance}
\label{sec:concept}

Governance can be broadly defined as the management and enforcement of control. This is inspired by DuPont's suggestion to interpret it in the realm of decentralized autonomous organizations (DAOs) \cite{Voshmgir.2017} as "\textit{...a mechanism that sets institutional rules and incentives, or the strategic exercise of power...}"\cite{DuPont_2019}. Yet, it remains challenging to provide a more suitable or tailored definition without using a specific lens \cite{Hofman2021} or without focusing on a particular aspect of the system under investigation. This article examines the governance of control over a ledger-anchored DID at the ledger's application layer. Control over such a DID refers to the authority to update its ledger-anchored DDO. To gain control of a DID, an entity must be appointed as a DID controller by an already authorized entity, such as another DID controller. The latter, therefore, plays a key role in delegating control and determining who else is authorized to control the DID and to update, besides others, the governance policies in the DDO. Consequently, whoever has update power over the ledger-anchored DDO, governs the DID. 

Technically, a ledger-anchored DDO can be interpreted as a non-transferable identity artifact, like a soulbound token\cite{Buterin:2022}, that is recorded on a collectively operated ledger. Unlike soulbound tokens, which are referenced in an SC, owned by a single party, contain claims, and become immutable once minted, a ledger-anchored DDO primarily holds verification material, remains updatable, and can be jointly controlled by multiple entities. Co-control leads to the peculiarity that any DDO state update, including the governance policies, may require collective consent from some or all DID controllers. As such, governance of a ledger-anchored DID is defined as the framework of rules and processes by which multiple entities coordinate and collectively exercise control over an on-chain DDO. 

\begin{figure}[t!]
\centerline{\includegraphics[scale=0.83]{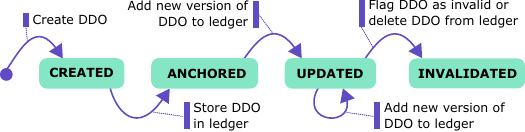}}
\caption{Lifecycle of a ledger-anchored DDO}
\label{fig:lifecycle}
\end{figure}

The lifecycle of a ledger-anchored DDO is described here with four states: CREATED, ANCHORED, UPDATED, and INVALIDATED. The creation deals with the generation of at least one asymmetric key pair, the DID and DDO. It happens ideally locally at the DID subject's premises, off-chain. The creation can also be taken care of by an entity different from the DID subject that acts as the DID subject's deputy in the digital realm. This step is followed by anchoring the DID in the VDR respectively persisting the DDO in a distributed ledger so that the DID is linked to a single, unique, tamper-proof, and verifiable version of the DDO. If anything changes regarding the DID, such as keys are rotated for security purposes, the DDO is updated in the ledger to ensure that others become immediately aware of the latest version. If a DID subject ceases to exist and the DID is not needed anymore then the DDO is invalidated by deleting it from a redactable ledger or by flagging the DID as invalid. The lifecycle is illustrated in Figure \ref{fig:lifecycle}. 

After a DDO is created, the DID is governed by the entity creating it. As soon as the DDO gets anchored in a ledger, it becomes governable by other parties as the single source of truth has moved from the creator's premises to the collectively operated ledger. Now the ledger becomes the platform through which governance policies of the DDO are managed and enforced. In other words, the governance of a ledger-anchored DID kicks in as soon as the DDO is anchored in the ledger and needs to be updated. It can be updated for various reasons. For instance, keys are rotated or DID controllers are appointed. The invalidation of a ledger-anchored DDO is a special form of update and is as such subsumed under a DDO update. To gain a clearer understanding of what it means to grant a DDO update as a governance act, it is necessary to discuss how governance policies are represented in the DDO, interpreted, and enforced through a ledger-based DID method.

\subsection{Policy Definition}
Only a DID controller has the right to perform or initiate a DDO update. The extent to which a DDO can be updated is neither specified nor specifiable in a standard-conform manner, resulting in most DID methods granting a DID controller full update power, as pointed out in Section \ref{sec:related}. The governance policy assigning an entity the DID controller role in the DDO materializes in various forms.   

\begin{figure}[t!]
\centerline{\includegraphics[scale=0.86]{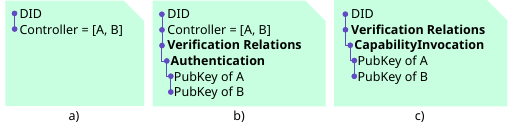}}
\caption{Examples of governance policies in the DDO with similar effect: a) List of DID controllers, b) List of DID controllers and their verification methods used for the DDO update, and c) List of the DID controllers verification methods.}
\label{fig:document}
\end{figure}

One of three proposed ways by the DID specification to declare a entity to be a DID controller is by enlisting its DID in the \texttt{controller} property on the highest level of the DDO. For enlisted DID controllers (except the DID subject) there is no need to have their public key material being present in the DID subject's DDO to properly authenticate themselves for the update as long as the required public key material is present in their own DDOs. Figure \ref{fig:document} illustrates the variants without (a) and with (b) the DID controller's public key material declared in the DID subject's DDO. If the public key material of a DID controller in the DID subject's DDO differs from that in the DID controller's DDO, the DID method must define which one takes precedence. The current DID specification lacks any clarification on how to deal with this type of ambiguity.  

DID controllers can also be authorized in the DDO's \texttt{capabilityInvocation} section. It lists verification methods for entities that are permitted to invoke a certain capability on behalf of the DID subject. The capability is, however, determined by the DID method. For example, it can be interpreted as the ability to access a restricted REST end point. If it is interpreted as the ability to update the DDO, then the DID controller's public key material can be listed in this section, rendering the \texttt{controller} property obsolete. Figure \ref{fig:document} c) illustrates this variant. A third option is closely related, as it is possible, instead, to enlist DID controller's verification methods in the \texttt{authentication} section if the DID method interprets these as entity's with DDO update authority. The difference lies in the location of the public key material needed to conduct the update. DID controllers enlisted in the aforementioned sections must have their public key material for the DDO update in the DID subject's DDO while the ones enlisted in the \texttt{controller} property are free to do so.    

All variants permit multiple DID controllers to be declared. The DID specification describes two potential governance scenarios if a DID is under control of multiple entities. With \textit{independent control}, each DID controller is empowered to invoke a DDO update without involving nor notifying other DID controllers. There is no coordination needed. With \textit{group control}, a DDO update needs to be coordinated among the DID controllers but is finally submitted by a single DID controller that represents the group. A hybrid model where each listed DID controller has independent control and some or all of them are managed by groups of entities is also possible. How the members of the group coordinate each other is out of the DID specification's scope.

After all, it lies in the hands of the ledger-based DID method to prescribe how a governance policy must look like in the DDO. A ledger-based DID method is also permitted to completely forgo governance policies in the DDO and instead leave it to the ledger's access control and its policies to decide about DDO updates. This, however, requires the ledger's access control layer to become aware of the concept of DIDs. But enlisting governance policies in the DDO and enforcing them through ledger- or VDR-specific mechanisms offers advantages such as transparency, interoperability, and auditability. By publishing governance policies in a standardized form, interoperability is facilitated, as different ledgers can consistently interpret and enforce these in a general applicable form. It lays the foundation for DID governance to become assured and transferable across different ledgers, ledger technologies, and non-ledger VDR solutions. In addition, explicit update authorizations provide a clear history of permissions, which is important for auditability and regulatory compliance. 

\subsection{Policy Enforcement}

One of the advantages of using ledger technology as the underlying VDR technology for DIDs is that governance policies of DIDs can be enforced in a decentralized and transparent fashion, without having to rely on a single central authority. It means every participant that is involved in the consensus process has the chance to verify whether an entity is authorized to update the DDO. This can be achieved in different ways. A ledger can support DIDs natively and come with built-in transaction types and logic to enforce authorizations, such as with multi-signature wallets\cite{Han.2021}. Hyperledger Indy allows a legal Indy entity with write permissions to write objects on the ledger. These can contain all DID-related information of a DDO. Hence, writing an object with such information corresponds logically to anchoring a DDO in the ledger. The legal Indy entity becomes hereby the sole DID controller. Every transaction initiating an update of such an object must then be signed by this legal Indy entity to become valid. Alternatively, SCs in a general-purpose ledger can be used to anchor DDOs. Chain codes to store and manage DDOs are publicly available for, e.g., Hyperledger Fabric\footnote{https://www.lfdecentralizedtrust.org/projects/fabric} or Ethereum. Yet, most ledgers supporting DIDs or SCs implementing a VDR assume single-entity DID control, while those allowing multiple controllers lack ledger-supported coordination for DDO updates.

\subsection{Potential Enhancements}

The following subsection discusses various enhancements to the DID concept, including multiple mechanisms for appointing DID controllers, different approaches to define roles and authorizations, and alternative ways to enforce governance policies.

\subsubsection{Fine-granular Authorizations and Roles}

As pointed out in \cite{vulnerability_did_doc_update, modid, Mazzocca.2025}, delegating control to an entity that is different to the DID subject comes with a security risk. The support for multiple DID controllers per DID increases the risk even further since with independent control they share the same update power. To minimize the potential damage a malicious DID controller could cause, their authorization should be strictly limited to essential permissions, adhering to the principle of least privilege. The ability to define and assign fine-grained authorizations within a DDO could, in certain scenarios, justify granting update permissions to entities that have so far acted solely as DID delegates. For instance, an entity becomes a DID delegate by adding its verification method to a DID subject's DDO. That makes the DID delegate the "owner" of its verification method in the DID subject's DDO. However, a key rotation for a DID delegate must be conducted by a DID controller on behalf of the DID delegate because the latter has no update authorization. With fine-granular authorizations, however, each DID delegate could be granted permission to update only its "own" artifacts in the DID subject's DDO. Hence, they would become DID controllers with minimal update power.

The degree to which update authorizations can be fine-tuned is limited by the set of resource elements in a DDO. At its core, a DDO contains verification methods of the DID subject, DID delegates, and DID controllers. As mentioned above, it is also declared for what purpose a enlisted verification method can be used for. In addition, the DDO may contain other resource elements such as the list of DID controllers, a list of alternative identifiers of the DID subject, and a list of technical contact points that the DID subject or associated entities are offering. The DID subject's identifier in the DDO is immutable over the lifetime of a DDO and is as such not changeable. Meta information such as the DDO's status (active or invalid) can be interpreted as another changeable resource element even though not being part of the DDO. All these resource elements, except the DID subject's identifier, represent the logical artifacts that can be affected by a DDO update. The operations add, update, and remove describe how the artifact of a DDO gets affected. The set of potential authorizations for a DDO update contains all combinations of what can be affected, how, and by whom. 

Fine-granular authorizations comprising the authorized subject and the authorized operation could be defined for each resource element in the DDO. Alternatively, roles with specific sets of fine-granular authorizations could be defined globally for a DID method in a single ledger-anchored document, e.g., using the access control policy description language XACML\cite{xaml} or ALFA\cite{alfa}. They could also be declared individually per DID, inline within the DDO in a dedicated section of its own. The assignment of roles to entities could either be done on the document level of the DDO or per resource element, with reference to the global or DID-specific role definitions.  

\subsubsection{Delegation with Tokens and VCs}

The ledger-recorded DDO is the single source of truth for governance policies unless the ledger's access control layer enforces its own policies. Instead, the DDO could be interpreted as a single source of truth for a list of trustful entities, here named \textit{DID trustees}, that are permitted to equip entities with update authorizations off-chain, without enlisting them explicitly in the DDO. The DID trusties, e.g., as enlisted in a \texttt{trustees} property, could issue a signed non-opaque bearer token for a DDO update off-chain directly to an entity, e.g., as a JSON Web Token\cite{rfc7519}. The possession of the token authorizes the entity towards the ledger to invoke a DDO update once, several times, or an unlimited number of times. It may also contain fine-granular authorizations. The authenticity and integrity of the token is verifiable by the ledger because a) a DID trustee's DDO is anchored in the ledger as well and b) its DID is explicitly enlisted in the DID subject's DDO as a trust anchor. This approach offers greater flexibility because not all update-authorized entities need to be enlisted in the DID subject's DDO as long as the token was issued by a DID trustee. If an authorization token should be bound to a particular entity, then a verifiable credential (VC) in a format of choice\cite{young.2021} may be used instead. With a VC, the invoker of the DDO update needs to present the signed permission to the ledger but it also has to prove to be in control of the ledger-anchored DID the VC is cryptographically bound to. In this case, the invoker DID would be revealed during the DDO update. If the invoker possesses no ledger-anchored DID or the invoker's ledger-anchored DID should not be revealed, then an ephemeral non-ledger-anchored and randomly-created DID in form of a \texttt{did:key} \cite{didkey} or \texttt{did:peer} \cite{didpeer} could be used instead. Yet, DID trustees have to make sure that the entity the token or VC is issued to is entitled to update the DDO on behalf of the DID subject. 

\subsubsection{Programmable Coordination}

With group control, the DID specification permits entities to coordinate a DDO update among each other and submit it finally under the roof of a common and enlisted DID controller. If the coordination happens off-chain, resulting in a potentially multi-party-signed DDO update, then it lies in the hands of the group members to arrange their coordination securely and trustfully using techniques such as Multi-party Computation\cite{Zhao.2019}, Threshold Signature Schemes\cite{Shoup.2000}, or similar. Coordination can also be conducted transparently through the ledger, like the collaborative administration model for access control policies proposed in \cite{Crass.2022}. With programmable and decentrally executable logic in the form of SCs, all voting schemes \cite{Ding.2023} used in on-chain governance models for DAOs could serve as potential coordination mechanisms among accredited entities. An entity authorized to do so would announce a DDO update proposal transparently via an SC. Entities with dedicated voting rights for the DDO are then capable to approve or reject a DDO update proposal via the same SC. Examples of potential voting mechanism are weighted, n-out-of-m, unanimous and delegated voting, with or without time restrictions, anonymously or non-anonymously, and with either a policy-based authorization proof to vote or by means of tokens or VCs. 

\subsubsection{Group-specific Coordination}

A coordination mechanism may apply globally to the DID method or, if namable in the DDO, only in the scope of a single DID.  If the DID controllers of a single DID can be divided into groups by tagging each registered controller with a group name, it becomes possible to apply group-specific coordination mechanisms for that DID. For example, a group of DID controllers with full update authority might need to agree unanimously on a critical governance-related DDO update while it is sufficient for another group with the authority to add an alternative DID subject identifier to reach an n-out-of-m consensus on the DDO update. 

\subsubsection{Adaptable Coordination}

Another advantage of explicitly declared coordination mechanisms is that it becomes a resource element of the DDO and can therefore be changed through a DDO update. The governance of a DID respectively becomes hereby modifiable by the DID controllers themselves, making the governance of a DID adaptable during a its lifetime.   

\subsubsection{Imperatively Specified Coordination}

The coordination processes for a DID update may also be specifiable in an imperative manner in the DDO or referenceable by the DDO, using a modeling language such as Business Process Model and Notation (BPMN) \cite{bpmn}, Petri Nets, or Decision Model and Notation (DMN)\cite{dmn}. As first investigated for the modeling of the blockchain governance for permissionless ledgers in \cite{Shiva.2021}, this can help to assess a governance model's fairness and other characteristics before deployment. The governance models would then be translatable into executable ledger-specific chain code \cite{Bodorik.2023}. The advantage of using a imperative specification with a standard notation is that it enables the coordination process, as an essential element of governance, to be interpretable and as such processable across different DID methods.

\section{Implementation}
\label{sec:implementation}

To demonstrate the technical feasibility of the enhancements discussed, a prototypical DID-conform identifier management system was implemented as multiple SCs for Ethereum. A single SC acts as the VDR for DDOs, allowing the anchoring and retrieval of DDOs and the announcement of DDO update proposals to initiate an update coordination process. Each update coordination mechanism is encoded as a separate SC and implements a common interface to be transparently replaceable during operation. The prototype relies on Ethereum's built-in account mechanism rather than using DIDs, without affecting the transferability of the solution because DID-conform DDOs can be generated on demand for retrieval purposes. Consequently, DID subjects, delegates and controllers are each identified by their ETH address instead of a VDR-agnostic DID.  

A DDO is parameterized with a DID-specific governance configuration. It contains details such as the SC's addresses of the update coordination mechanism(s), optional expiration dates, DID controllers and trustees, and more. Each instance of an update coordination mechanism represents a voting for a proposed DDO update. It can last until a condition specific to the voting type is fulfilled or a time limit is reached, depending on the governance configuration. Chainlink Automation\footnote{https://docs.chain.link/chainlink-automation} is applied as an external timekeeper for time-limited coordination. The following update coordination mechanisms were implemented: unanimity, n-out-of-m, and weighted majority. The prototype supports three types of proofs for the authorization to announce a DDO update and to approve or reject it. An entity is authorized if it is assigned the role of a DID controller in the DDO or because it is in possession of either an authorization token or VC as issued by a DDO-declared DID trustee. For the VC, the entity needs, in addition, to prove to have control over the DID the VC is cryptographically bound to. 

DID controllers of a DDO can be assigned to one of four groups, each utilizing a potentially different update coordination process and being granted a distinct set of update authorizations. Members of group A with the least privileges are allowed to update functional policies of the DDO that do not affect the DDO's governance, such as to appoint DID delegates. Members of group B are, in addition, permitted to update the governance policies of their own group. Members of group C can, in addition, partially delegate control by adding groups with the same authorizations as group A. Group D members can update the DDO without restrictions.    

In summary, the prototype supports customizable groups with fine-granular authorizations for partial DDO updates, multiple implemented coordination mechanisms with or without expiration time, group-specific coordination, and the ability of the governance to be updated during the DID's lifecycle. The code of the prototype can be obtained here: \url{https://github.com/TU-Berlin-SNET/DID-IDM}.

\section{Conclusion and Future Work}
\label{sec:conclusion}

Allowing multiple entities to commonly control a ledger-anchored DID is a powerful feature. It extends the range of applications for DIDs to include use cases in which complex custodial, ownership, legal, and administrative relationships among natural persons, legal entities, and non-human entities need to be technically enforceable through the governance of the associated DID-based digital identities. The discussion about today's deployed update governance models and on the framework set by the current DID specification, and the exploration of the possible enhancements and extensions of the framework highlight the crucial role governance policies in the DDO play for the flexibility and applicability of a DID method and its interoperability with other DID methods. Equally important is how these policies are technically enforced by the underlying VDR to ensure transparency, auditability, and consequently the trustworthiness of a ledger-based realization of a DID method. The proof of concept based on SCs demonstrates, as a first of its kind, the technical feasibility to realize a DID-conform and decentralized identifier management system where a DID is governed by multiple entities and where - at the same time - the governance is fully customizable and transparently replaceable during the DID's lifetime.  

The governance of a DID has primarily been studied under the assumption that a predefined set of governance mechanisms applies uniformly to all DIDs of a DID method. To enable DID-specific governance and ensure interoperability, including cross-ledger and cross-VDR governance of DIDs, it must be possible to define governance policies in a standardized and universally applicable manner within the DDO or a referenced document. This would enable a DID to be governed by entities that have their DIDs anchored in different ledgers and/or VDR technologies. Yet, it still needs to be investigated how the various DID method-specific governance policies known today can be sufficiently generalized so that they can not only be clearly declared in the DDO but also be technically enforced across different decentralized VDRs in the form of distributed ledgers or decentralized file systems such as the interplanetary file system IPFS.

\section{Acknowledgements}
We would like to thank Witold Jerzy Jermakowicz, Baki Berkay Uzel, and Cem Deniz Dolanmaz from the Technische Universit\"at Berlin for their valuable advice and support. This work was funded by the Federal Ministry of Education and Research (BMBF) in Germany under the grant number 16KIS2251 of the SUSTAINET\_guarDian project.

\bibliographystyle{IEEEtran}  
\bibliography{references}  

\begin{thebibliography}{10}
\providecommand{\url}[1]{#1}
\csname url@samestyle\endcsname
\providecommand{\newblock}{\relax}
\providecommand{\bibinfo}[2]{#2}
\providecommand{\BIBentrySTDinterwordspacing}{\spaceskip=0pt\relax}
\providecommand{\BIBentryALTinterwordstretchfactor}{4}
\providecommand{\BIBentryALTinterwordspacing}{\spaceskip=\fontdimen2\font plus
\BIBentryALTinterwordstretchfactor\fontdimen3\font minus \fontdimen4\font\relax}
\providecommand{\BIBforeignlanguage}[2]{{%
\expandafter\ifx\csname l@#1\endcsname\relax
\typeout{** WARNING: IEEEtran.bst: No hyphenation pattern has been}%
\typeout{** loaded for the language `#1'. Using the pattern for}%
\typeout{** the default language instead.}%
\else
\language=\csname l@#1\endcsname
\fi
#2}}
\providecommand{\BIBdecl}{\relax}
\BIBdecl

\bibitem{Toth.2019}
K.~C. Toth and A.~Anderson-Priddy, ``{Self-Sovereign Digital Identity: A Paradigm Shift for Identity},'' \emph{IEEE Security {\&} Privacy}, vol.~17, no.~3, pp. 17--27, 2019.

\bibitem{Avellaneda.2019}
O.~Avellaneda, A.~Bachmann, A.~Barbir, J.~Brenan, P.~Dingle, K.~H. Duffy, E.~Maler, D.~Reed, and M.~Sporny, ``{Decentralized Identity: Where Did It Come From and Where Is It Going?}'' \emph{IEEE Communications Standards Magazine}, vol.~3, no.~4, pp. 10--13, 2019.

\bibitem{WorldWideWebConsortium.822021}
{World Wide Web Consortium (W3C)}, ``{Decentralized Identifiers v1.0},'' \url{https://www.w3.org/TR/did-core/}, 2022, accessed: 2025-05-13.

\bibitem{WorldWideWebConsortium.VC}
------, ``{Verifiable Credentials Data Model v2.0},'' \url{https://www.w3.org/TR/vc-data-model-2.0/}, 2025, accessed: 2025-05-13.

\bibitem{Allen.2016}
{Christopher Allen}, ``{The Path to Self-Sovereign Identity },'' \url{https://www.lifewithalacrity.com/article/the-path-to-self-soverereign-identity/}, 2016, accessed 2025-05-13.

\bibitem{did_use_cases}
{World Wide Web Consortium (W3C)}, ``{Use Cases and Requirements for Decentralized Identifiers},'' \url{https://www.w3.org/TR/did-use-cases/}, 2021, accessed: 2025-05-13.

\bibitem{Sandhu.1996}
R.~S. Sandhu, E.~J. Coyne, H.~L. Feinstein, and C.~E. Youman, ``Computer role-based access control models,'' \emph{Computer}, vol.~29, no.~2, p. 38 – 47, 1996.

\bibitem{vulnerability_did_doc_update}
M.-H. Rhie, K.-H. Kim, D.~Hwang, and K.-H. Kim, ``\BIBforeignlanguage{en}{Vulnerability {Analysis} of {DID} {Document}’s {Updating} {Process} in the {Decentralized} {Identifier} {Systems}},'' in \emph{\BIBforeignlanguage{en}{2021 {Int.} {Conf.} on {Information} {Networking} ({ICOIN})}}.\hskip 1em plus 0.5em minus 0.4em\relax Jeju Island, Korea (South): IEEE, Jan. 2021, pp. 517--520.

\bibitem{modid}
H.~Yang, R.~Song, B.~Chen, Y.~Song, and B.~Xiao, ``\BIBforeignlanguage{en}{{MoDID}: {Decentralized} {Identity} {Management} for {Multiple} {Owners}},'' in \emph{\BIBforeignlanguage{en}{{ICC} 2024 - {IEEE} {Int.} {Conf.} on {Communications}}}.\hskip 1em plus 0.5em minus 0.4em\relax Denver, CO, USA: IEEE, Jun. 2024, pp. 2022--2027.

\bibitem{dad_3_rs}
S.~M. Smith and V.~Gupta, ``{Decentralized Autonomic Data (DAD) and the three R's of Key Management},'' \url{https://github.com/SmithSamuelM/Papers/blob/master/whitepapers/DecentralizedAutonomicData.md}, 2018, accessed 2025-05-13.

\bibitem{didmethods}
{World Wide Web Consortium (W3C)}, ``{DID Methods},'' \url{https://www.w3.org/TR/did-extensions-methods/}, 2025, accessed 2025-05-13.

\bibitem{Hoops.2023}
F.~Hoops, A.~Muhle, F.~Matthes, and C.~Meinel, ``{A Taxonomy of Decentralized Identifier Methods for Practitioners},'' in \emph{{2023 IEEE Int. Conf. on Decentralized Applications and Infrastructures (DAPPS)}}.\hskip 1em plus 0.5em minus 0.4em\relax IEEE Computer Society, 2023, pp. 57--65.

\bibitem{Bistarelli.2023}
S.~Bistarelli, F.~Micheli, and F.~Santini, ``{A Survey on Decentralized Identifier Methods for Self Sovereign Identity},'' in \emph{{Proceedings of the Italian Conf. on Cyber Security (ITASEC 2023)}}, 2023.

\bibitem{didethr}
{Veramo core team}, ``{ETHR DID Method Specification},'' \url{https://github.com/decentralized-identity/ethr-did-resolver/blob/master/doc/did-method-spec.md}, 2022, accessed 2025-05-13.

\bibitem{didbnb}
{ONTology Tech}, ``{Binance DID Method Specification},'' \url{https://github.com/ontology-tech/DID-method-specs/blob/master/did-bnb/DID-Method-bnb.md}, 2020, accessed 2025-05-13.

\bibitem{didindy}
{Stephen Curran, Paul Bastian, Daniel Hardman, Char Howland, Christian Bormann, Dominic Wörner, Daniel Bluhm, Kyle Den Hartog, Artem Ivanov, Renata Toktar, Alexander Shcherbakov}, ``{Indy DID Method},'' \url{https://hyperledger.github.io/indy-did-method/}, accessed 2025-05-13.

\bibitem{didsov}
{World Wide Web Consortium (W3C)}, ``{Sovrin DID Method Specification},'' \url{https://sovrin-foundation.github.io/sovrin/spec/did-method-spec-template.html}, 2025, accessed 2025-05-13.

\bibitem{didalgo}
{Cosimo Bassi, Joe Polny, Bruno Martins}, ``{did-algo SPEC},'' \url{https://github.com/algorandfoundation/did-algo/blob/main/SPEC.md}, 2024, accessed 2025-05-13.

\bibitem{didsol}
{Riedel, Martin and Kelleher, Daniel}, ``{The did:sol Method v3.0},'' \url{https://g.identity.com/sol-did/}, 2023, accessed 2025-05-13.

\bibitem{didebsi}
{European Blockchain Services Infrastructure (EBSI)}, ``{DID Method for Legal Entities},'' \url{https://hub.ebsi.eu/vc-framework/did/legal-entities}, 2025, accessed 2025-05-13.

\bibitem{didfactom}
{Valentin Ganev, Peter Asenov, Niels Klomp, Carl DiClementi, Sam Barnes}, ``{Factom Decentralized Identifiers (DID)},'' \url{https://github.com/factom-protocol/FIS/blob/master/FIS/DID.md}, 2019, accessed 2025-05-13.

\bibitem{didiota}
{IOTA Foundation}, ``{IOTA DID Method Specification v2.0},'' \url{https://docs.iota.org/references/iota-identity/iota-did-method-spec}, 2024, accessed 2025-05-13.

\bibitem{Mazzocca.2025}
C.~Mazzocca, A.~Acar, S.~Uluagac, R.~Montanari, P.~Bellavista, and M.~Conti, ``{A Survey on Decentralized Identifiers and Verifiable Credentials},'' \emph{IEEE Communications Surveys \& Tutorials}, pp. 1--1, 2025.

\bibitem{Voshmgir.2017}
V.~Shermin, ``{Disrupting governance with blockchains and smart contracts},'' \emph{Strategic Change}, vol.~26, no.~5, pp. 499--509, 2017.

\bibitem{DuPont_2019}
Q.~DuPont, \emph{Cryptocurrencies and blockchains}.\hskip 1em plus 0.5em minus 0.4em\relax Polity, 2019.

\bibitem{Hofman2021}
D.~Hofman, Q.~DuPont, A.~Walch, and I.~Beschastnikh, \emph{{Blockchain Governance: De Facto (x)or Designed?}}\hskip 1em plus 0.5em minus 0.4em\relax Cham: Springer International Publishing, 2021, pp. 21--33.

\bibitem{Buterin:2022}
P.~Ohlhaver, E.~G. Weyl, and V.~Buterin, ``{Decentralized Society: Finding Web3's Soul},'' \url{https://ssrn.com/abstract=4105763}, Tech. Rep., 2022, accessed 2025-05-13.

\bibitem{Han.2021}
J.~Han, M.~Song, H.~Eom, and Y.~Son, ``{An Efficient Multi-signature Wallet in Blockchain Using Bloom Filter},'' in \emph{Proceedings of the 36th Annual ACM Symposium on Applied Computing}, ser. SAC '21.\hskip 1em plus 0.5em minus 0.4em\relax New York, NY, USA: Association for Computing Machinery, 2021, p. 273–281.

\bibitem{xaml}
{OASIS}, ``{eXtensible Access Control Markup Language (XACML) Version 3.0},'' \url{http://docs.oasis-open.org/xacml/3.0/xacml-3.0-core-spec-os-en.html}, 2013, accessed 2025-05-13.

\bibitem{alfa}
------, ``{Abbreviated Language for Authorization (ALFA)},'' \url{https://alfa.guide/}, accessed 2025-04-14.

\bibitem{rfc7519}
M.~Jones, J.~Bradley, and N.~Sakimura, ``{JSON Web Token (JWT)},'' Internet Requests for Comments, RFC Editor, RFC 7519, 05 2015.

\bibitem{young.2021}
{Kaliya Young}, ``{Verifiable Credentials Flavors Explained},'' \url{https://www.lfph.io/wp-content/uploads/2021/02/Verifiable-Credentials-Flavors-Explained.pdf}, accessed: 2025-05-13.

\bibitem{didkey}
{World Wide Web Consortium (W3C)}, ``{The did:key Method v0.7 (unofficial draft)},'' \url{https://w3c-ccg.github.io/did-key-spec/}, 2025, accessed 2025-05-13.

\bibitem{didpeer}
{Decentralized Identity Foundation}, ``{Peer DID Method Specification},'' \url{https://identity.foundation/peer-did-method-spec/}, accessed 2025-05-13.

\bibitem{Zhao.2019}
C.~Zhao, S.~Zhao, M.~Zhao, Z.~Chen, C.-Z. Gao, H.~Li, and Y.~an~Tan, ``{Secure Multi-Party Computation: Theory, practice and applications},'' \emph{Information Sciences}, vol. 476, pp. 357--372, 2019.

\bibitem{Shoup.2000}
V.~Shoup, ``{Practical Threshold Signatures},'' in \emph{Advances in Cryptology --- EUROCRYPT 2000}, B.~Preneel, Ed.\hskip 1em plus 0.5em minus 0.4em\relax Springer Berlin Heidelberg, 2000, pp. 207--220.

\bibitem{Crass.2022}
S.~Craß, A.~Lackner, N.~Begic, S.~A.~M. Mirhosseini, and N.~Kirchmayr, ``{Collaborative Administration of Role-Based Access Control in Smart Contracts},'' in \emph{2022 4th Conf. on Blockchain Research \& Applications for Innovative Networks and Services (BRAINS)}, 2022, pp. 87--94.

\bibitem{Ding.2023}
Q.~Ding, W.~Xu, Z.~Wang, and D.~Lee, ``{Voting Schemes in DAO Governance},'' \emph{World Scientific Annual Review of Fintech}, 2023.

\bibitem{bpmn}
{Object Management Group}, ``{Business Process Model and Notation},'' \url{https://www.bpmn.org/}, accessed 2025-05-13.

\bibitem{dmn}
------, ``{Decision Model and Notation™ (DMN™)},'' \url{https://www.omg.org/dmn/}, accessed 2025-05-13.

\bibitem{Shiva.2021}
S.~Jairam, J.~Gordijn, I.~Da, S.~Torres, F.~Kaya, and M.~Makkes, ``{A Decentralized Fair Governance Model for Permissionless Blockchain Systems},'' in \emph{CEUR Proceeding of the Workshop of Value Modelling and Business Ontologies}, vol. 2835, 2021.

\bibitem{Bodorik.2023}
P.~Bodorik, C.~G. Liu, and D.~Jutla, ``{TABS: Transforming automatically BPMN models into blockchain smart contracts},'' \emph{Blockchain: Research and Applications}, vol.~4, no.~1, p. 100115, 2023.

\end{thebibliography}

\end{document}